\documentclass[11pt,twoside]{article}
\usepackage{asp2009}
\usepackage{epsf}
\usepackage{lscape}
\usepackage{natbib}
\usepackage{graphicx}

\begin{document}
\bibliographystyle{asp2010}

\title{Interacting compact binaries: modeling mass transfer in eccentric
systems}

\author{Ross P. Church$^1$, Melvyn B. Davies$^1$, Alexey Bobrick$^1$, and Christopher A. Tout$^2$
  \affil{$^1$Department of Astronomy and Theoretical Physics, Lund Observatory,
  Box 43, SE221 00 Lund, Sweden\\
  $^2$Institute of Astronomy, The Observatories, University of Cambridge,
  Madingley Road, Cambridge.  CB3 0HA.  UK.}
  }

\begin{abstract}
We discuss mass transfer in eccentric binaries containing a white dwarf and a
neutron star (WD--NS binaries).  We show that such binaries are produced from
field binaries following a series of mass transfer episodes that allow the white
dwarf to form before the neutron star.  We predict the orbital properties of
binaries similar to the observed WD--NS binary J1141+6545, and show that they
will undergo episodic mass transfer from the white dwarf to the neutron star.
Furthermore, we describe oil-on-water, a two-phase SPH formalism that we have
developed in order to model mass transfer in such binaries.
\end{abstract}


\section{Introduction}
Interacting compact binaries are responsible for a wide range of astronomically
interacting events.  Accreting white dwarfs (WDs) are believed to be the
progenitors of type IA supernovae.  Cataclysmic variables also occur
when white dwarfs accrete matter but under different conditions.
Higher-mass accretors -- neutron stars (NSs) or black holes (BHs) form X-ray
binaries.  A recent paper \citep{metzger11} suggests that the tidal
disruption of a WD by either a NS or a BH may lead to an accretion flow that is
dominated by nuclear burning, with a potentially unique nucleosynthetic
signature.

Here we consider the problem of modeling the onset of mass transfer in an eccentric binary
containing a white dwarf and a neutron star, a potential progenitor of
such a nuclear-dominated accretion flow.  We discuss the formation of such
binaries and show that when they come into contact they are expected to undergo
episodic mass transfer.  Finally we outline the oil-on-water SPH technique that
we have developed to study these binaries.

\section{Formation of eccentric WD--NS binaries}

There are two clear detections of eccentric binaries containing a white dwarf
and a neutron star; B2303+46 \citep{vanK99} and J1141-6545 \citep{Kaspi00}.
One might expect that such binaries would not exist, as the neutron
star, having a more massive and hence rapidly evolving progenitor, should form
first.  Tidal circularisation during the giant phase of the white dwarf's
progenitor would then remove any eccentricity imparted by the supernova.
Therefore, to produce such a binary, the white dwarf must form first.

A formation mechanism was outlined by \citet{pZ99} and
developed by \citet{Tauris00}, \citet{Davies02} and \citet{church06}.  In it,
the binary initially contains two main-sequence stars of similar, but not
identical, mass, and has an orbital separation of around $100\,R_{\odot}$.  
The initially most massive star fills its Roche Lobe in the Hertzsprung gap and
transfers mass stably to its companion.  Subsequent phases of mass
transfer can follow several pathways, but in all cases the initially most
massive star fills its Roche lobe again, transfers the majority of its mass
to its companion and finally becomes a white dwarf.  The companion
star is now massive enough to evolve a Chandrasekhar-mass degenerate core, which
collapses leading to a supernova and, if the resulting binary is
bound, the formation of an eccentric binary.  We refer to such systems as WD--NS
binaries.

\citet{church06} predict the binaries 
similar to J1141-6545 form when a fourth episode of mass transfer
occurs after the white dwarf has formed.  This mass transfer is inevitably from
the most massive star at this point (the progenitor of the neutron star) and
hence causes the orbit to shrink, producing a tighter binary after the
supernova.  Such close binaries merge in less than the Hubble time and hence are
of interest to us.  

\begin{figure}
\includegraphics[width=\columnwidth]{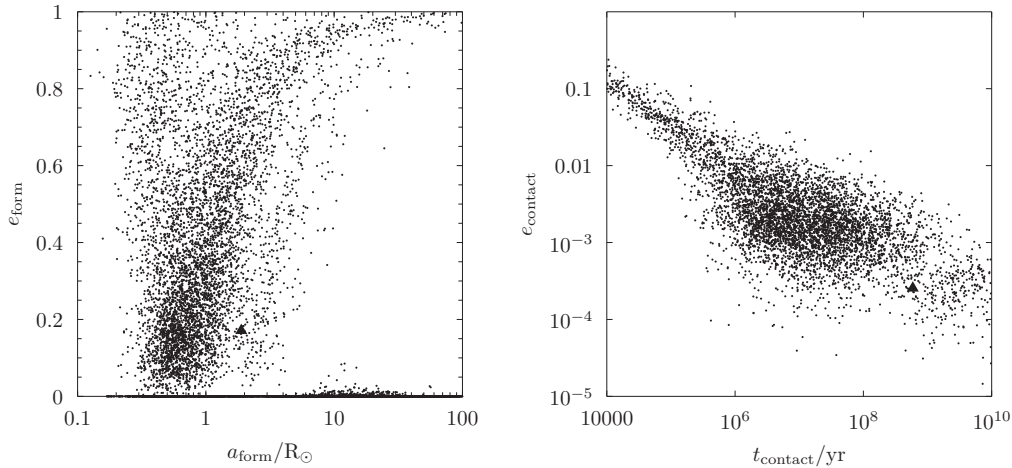}
\caption{Left panel: The eccentricities of our simulated sample of WD--NS
binaries as a function of their semi-major axis, both at formation time,
i.e.~just after the supernova.  The filled triangle represents J1141+6545.
Right panel: Eccentricities of WD--NS binaries at the point when the white dwarf
fills its Roche lobe, as a function of the time taken for them to spiral
together.}
\label{fig:econtact}
\end{figure}

Following the formation of the binary, the emission of gravitational waves
causes its orbit to shrink and circularise \citep{peters64}.  In the fully
circularised case, mass transfer from the white dwarf begins once it fills
its Roche lobe.  However, these binaries still have a residual eccentricity at
this point.  In Figure~\ref{fig:econtact} we plot the orbital properties for a
sample of WD--NS binaries that form via a pathway similar to that inferred for
J1141-6545, as computed by BSE \citep{hurley02}.  It can be seen that our
simulations produce binaries with orbits similar to that of J1141-6545.  
The residual eccentricity when these binaries come into contact is typically
between $10^{-2}$ and $10^{-4}$.  As the surface pressure scale height of white
dwarfs is very small these eccentricities are large enough for episodic mass
transfer to occur.

\section{Using smoothed-particle hydrodynamics in eccentric binaries}

In order to model episodic mass transfer in eccentric WD--NS binaries we have
developed an extension to SPH which we call oil-on-water.  SPH is a Lagrangian
particle-based method for solving the equations of hydrodynamics
\citep{lucy77,gingold77}.  The implementation that we use follows
\citet{benz90}.  In vanilla SPH, one requires a very high mass resolution to
model mass transfer between stars.  For example, in a cataclysmic variable 
the mass transfer rate is $10^{-9}$ to $10^{-8}\,M_{\odot}\,{\rm yr^{-1}}$
\citep{patterson84};  given a
typical orbital period of $1\,{\rm day}$ the mass transferred per
orbit is one part in $10^{12}$ of the total mass.  For studies of continuous
mass transfer in circular binaries this is not a problem, as the flow is steady
and hence can be considered over a large number of orbits.  In the case of
episodic mass transfer that switches on and off each orbit, however, a new
approach is required.

To avoid this problem we split the star up into two separate phases.  Heavy
``water'' particles form the body of the star and account for essentially all of
its mass.  Light ``oil'' particles form an atmosphere that is transferred during
interactions between stars.  As the mass of the oil particles is much less than
that of the water particles numerical stability requires them not to come into
contact.  To prevent this from happening we introduce an additional force to
stop the oil particles sinking into the star.  This additional force, acting on
an oil particle of mass $m_i$, depends on the water particle number gradient $n$
as
\begin{equation}
F_{\rm n}=\beta\frac{GM_{\odot}m_i}{R_{\odot}}\nabla n ; \qquad \nabla n =
\sum_j \nabla W^{\rm ow}(r_{ij},h_{ij})
\end{equation}
where $\beta$ is an adjustable parameter.  We utilise a different kernel $W^{\rm
ow}$ with a break point at a smaller multiple of the smoothing length $h$ in
order to keep the oil layer closer to the surface of the star.  Full details of
the method are given in \citet{church09}.  It should be noted that the oil
particles are {\it not} test particles; they obey the full SPH equations when
interacting with one another, which lets us follow the processes inside the
accretor's Roche lobe.  

\begin{figure}
\includegraphics[height=.36\textwidth]{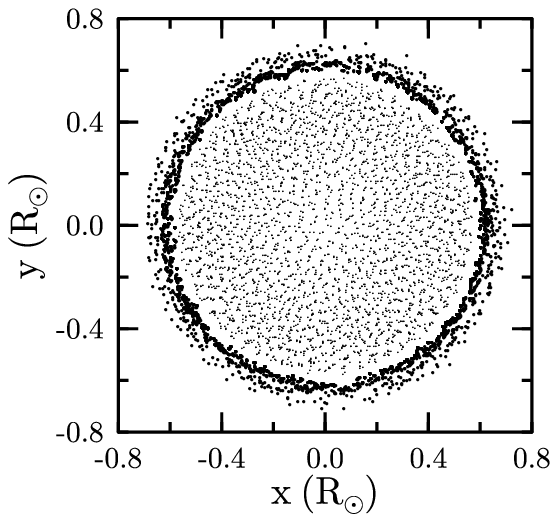}
\includegraphics[height=.36\textwidth]{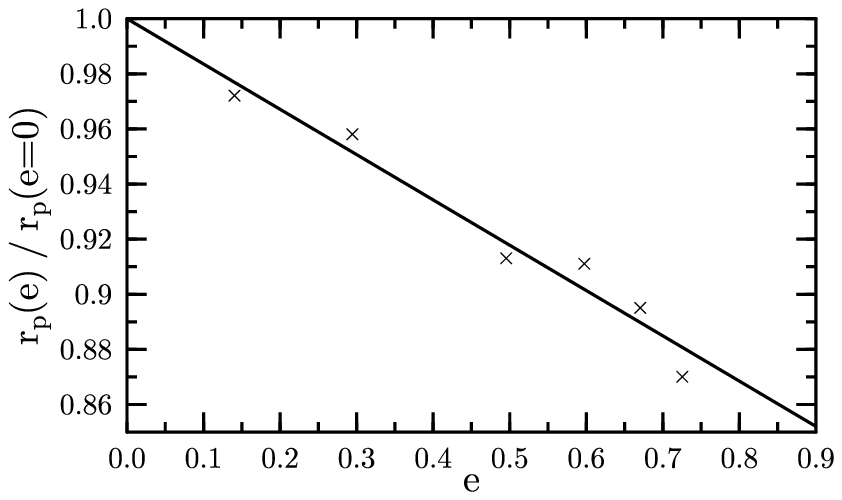}
\caption{Left panel: structure of an oil-on-water star.  Particles within $h$ of
the $x-y$ plane only are plotted.  The oil particles are plotted with larger
dots.  A clear atmosphere can be seen within the oil layer.  Right panel: the
ratio between the periastron separation required for mass transfer assuming that
the star must fill its Roche lobe at periastron and that measured from our
simulations, plotted as a function of eccentricity.  The minimum separation
required for mass transfer to take place is seen to decrease linearly with
increasing eccentricity.}
\label{fig:singlestar}
\end{figure}

To test the validity of the model 
we investigated mass transfer from a low-mass star to a white
dwarf in eccentric binaries.  For the body of the star we used 15\,390 water
particles with a total mass of $0.6\,M_{\odot}$.  The atmosphere was made up of
39\,691 oil particles with a mass of $10^{-14}\,M_{\odot}$ each.  This stellar model was
placed into a binary with a $1\,M_{\odot}$ white dwarf, modelled as a point
mass, and relaxed at apocentre.  The orbit was then made eccentric.  For each
eccentricity we varied the semi-major axis in order to find the pericentre
separation at which mass transfer started.  We found that mass was transferred
on each orbit shortly after pericentre passage, and that the semi-major axis
required for mass transfer decreased with increasing eccentricity, compared
with the value that would be implied by requiring the star to fill its Roche
lobe at pericentre (see Figure~\ref{fig:singlestar}).
This is in agreement with the results of \citet{sepinsky07}, who take an
analytical approach to determining the point of onset of mass transfer.

\acknowledgments 
RPC is funded by a Marie-Curie Intra-European Fellowship, Grant No.~252431 under
the European Comission's FP7 framework.  This work was supported by the Swedish
Research Council (grant 2008--4089).  The calculations we present were made
using computers purchased with the support of the Royal Physiographic Society of
Lund.

\bibliography{author} 

\end{document}